\documentclass{aastex63}

\submitjournal{AJ}

\shorttitle{Prediction model of XMM contamination}
\shortauthors{Kronberg et al.}

\begin{document}
	

\title{Prediction and understanding of soft proton contamination in XMM-Newton: a machine learning approach}

\correspondingauthor{Elena Kronberg}
\email{elena.kronberg@lmu.de}

\author[0000-0001-7741-682X]{Elena A. Kronberg}
\affiliation{Department of Earth and Environmental Sciences (Geophysics), University of Munich, Theresienstr. 41, Munich, 80333, Germany}

\author[0000-0002-9112-0184]{Fabio Gastaldello}
\affiliation{Instituto di Astrofisica Spaziale e Fisica Cosmica (INAF-IASF), Milano, via A. Corti 12, I-20133 Milano, Italy}

\author[0000-0002-1241-7570]{Stein Haaland}
\affiliation{Birkeland Centre for Space Science, University of Bergen, All\'{e}gaten 55, 5007 Bergen, Norway}
\affiliation{Max Planck Institute for Solar System Research, Justus-von-Liebig-Weg 3, G\"{o}ttingen, Germany}

\author[0000-0003-3689-4336]{Artem Smirnov}
\affiliation{German Research Centre for Geosciences, Albert-Einstein-Straße 42-46, Potsdam, 14473, Germany}
\affiliation{Geophysical Center of the Russian Academy of Sciences, Molodezhnaya St. 3, 119296 Moscow, Russia}

\author[0000-0001-9724-4009]{Max Berrendorf}
\affiliation{Institute of Informatics, University of Munich, Oettingenstraße 67,	Munich, 80538, Germany}

\author[0000-0003-0879-7328]{Simona Ghizzardi}
\affiliation{Instituto di Astrofisica Spaziale e Fisica Cosmica (INAF-IASF), Milano, via A. Corti 12, I-20133 Milano, Italy}

\author[0000-0001-6654-5378]{K. D. Kuntz}
\affiliation{Henry A. Rowland Department of Physics \& Astronomy, Johns Hopkins University, 3400 N. Charles Street, Baltimore, MD 21218, USA}

\author[0000-0003-4278-0482]{Nithin Sivadas}
\affiliation{Department of Electrical and Computer Engineering, Boston University, 8 Saint Mary's Street, Boston, MA 02134 USA}

\author[0000-0003-2079-5683]{Robert C. Allen}
\affiliation{Johns Hopkins University Applied Physics Lab, 11100 Johns Hopkins Rd, Laurel, MD 20723, USA}

\author[0000-0002-6038-1090]{Andrea Tiengo}
\affiliation{Scuola Universitaria Superiore IUSS Pavia, piazza della Vittoria 15, I-27100 Pavia, Italy}
\affiliation{Instituto di Astrofisica Spaziale e Fisica Cosmica (INAF-IASF), Milano, via A. Corti 12, I-20133 Milano, Italy}
\affiliation{INFN, Sezione di Pavia, via A. Bassi 6, I-27100 Pavia, Italy}

\author[0000-0002-7305-2579]{Raluca Ilie}
\affiliation{University of Illinois at Urbana-Champaign, 306 N. Wright Street, 5054 ECEB, Urbana, IL 61801, USA}

\author[0000-0001-5023-0427]{Yu Huang}
\affiliation{University of Illinois at Urbana-Champaign, 306 N. Wright Street, 5054 ECEB, Urbana, IL 61801, USA}

\author[0000-0002-8240-5559]{Lynn Kistler}
\affiliation{Space Science Center, University of New Hampshire, Morse Hall Rm 408, Durham, NH 03824, USA}





\begin{abstract}

One of the major and unfortunately unforeseen sources of background for the current generation of X-ray telescopes are few tens to hundreds of keV (soft) protons concentrated by the mirrors. One such telescope is the European Space Agency's (ESA) X-ray Multi-Mirror Mission (XMM-Newton). Its observing time lost due to background contamination is  about 40\%. This loss of observing time affects all the major broad science goals of this observatory, ranging from cosmology to astrophysics of neutron stars and black holes. The soft proton background could dramatically impact future large X-ray missions such as the ESA planned Athena mission\footnote{\url{http://www.the-athena-x-ray-observatory.eu/}}. Physical processes that trigger this background are still poorly understood. We use a Machine Learning (ML) approach to delineate related important parameters and to develop a model to predict the background contamination using 12 years of XMM observations. As predictors we use the location of satellite, solar and geomagnetic activity parameters. We revealed that the contamination is most strongly related to the distance in southern direction, $Z$, (XMM observations were in the southern hemisphere), the solar wind radial velocity and the location on the magnetospheric magnetic field lines. We derived simple empirical models for the first two individual predictors and an ML model which utilizes an ensemble of the predictors (Extra Trees Regressor) and gives better performance. Based on our analysis, future missions should minimize observations during  times  associated with high solar wind speed  and avoid closed magnetic field lines,  especially at  the dusk flank region in the southern hemisphere.

%


\end{abstract}

\keywords{X-ray telescopes (1825), X-ray detectors (1815), X-ray observatories (1819), Space plasmas (1544), Astronomy data modeling (1859), Astronomy data analysis (1858)}


\section{Introduction}\label{sec:intro}

X-ray telescopes are built to focus X-ray photons towards the detectors in the focal plane by a double low-angle scattering (grazing incidence) from concentric mirrors shells.  For the last two decades, with the advent of the modern X-ray observatories in orbit such as Chandra, see, e.g., \citet{Weisskopf02} and the X-ray Multi-Mirror Mission (XMM) Newton \citep{Jansen01}, it has been recognized that protons of energies in the range of tens of keV up to few MeVs, hereafter referred to as Soft Protons (SP), can scatter at low angles through the mirror shells and reach the focal plane, see, e.g., \citet{Fioretti16} and references therein. These protons, populating the 
interplanetary space and the Earth magnetosphere, can damage CCD detectors by delivering a non-ionising dose leading to a loss of spectral resolution. Their signal is indistinguishable from X-ray photons. Therefore, it can not be rejected and it produces an enhanced background. 

This phenomenon was 
discovered after the damaging of the Chandra/Advanced CCD Imaging Spectrometer (ACIS) front-illuminated (FI) CCDs 
during its first passage through the radiation belt \citep{Prigozhin00a}. Analysis of data of the 
calibration source showed that all the FI CCD chips had suffered some 
damage causing a significant increase in the Charge Transfer Inefficiency 
(CTI). CTI is caused by defects in the silicon lattice that can be created by 
the interaction with charged particles. These defects, or ``traps'', capture 
charges during their transfer to the read-out electronics, and release them 
at later times. Its effects on the detector performance are: 
position-dependent changes in the energy scale, loss of spectral resolution 
and loss of quantum efficiency \citep{Odell.ea:00,Prigozhin00b}. Therefore,
after less than two months of operation ACIS has been protected during 
radiation belt passages, by moving the detector out of the telescope focus 
\citep{Odell.ea:03}. 
The same procedure takes place during periods of enhanced particle flux, 
triggered either by the on board radiation monitor or by ground operations 
monitoring of various space weather probes \citep{Grant.ea:12}.

\begin{figure}[ht!]
	\includegraphics[angle=270, width=0.9\linewidth]{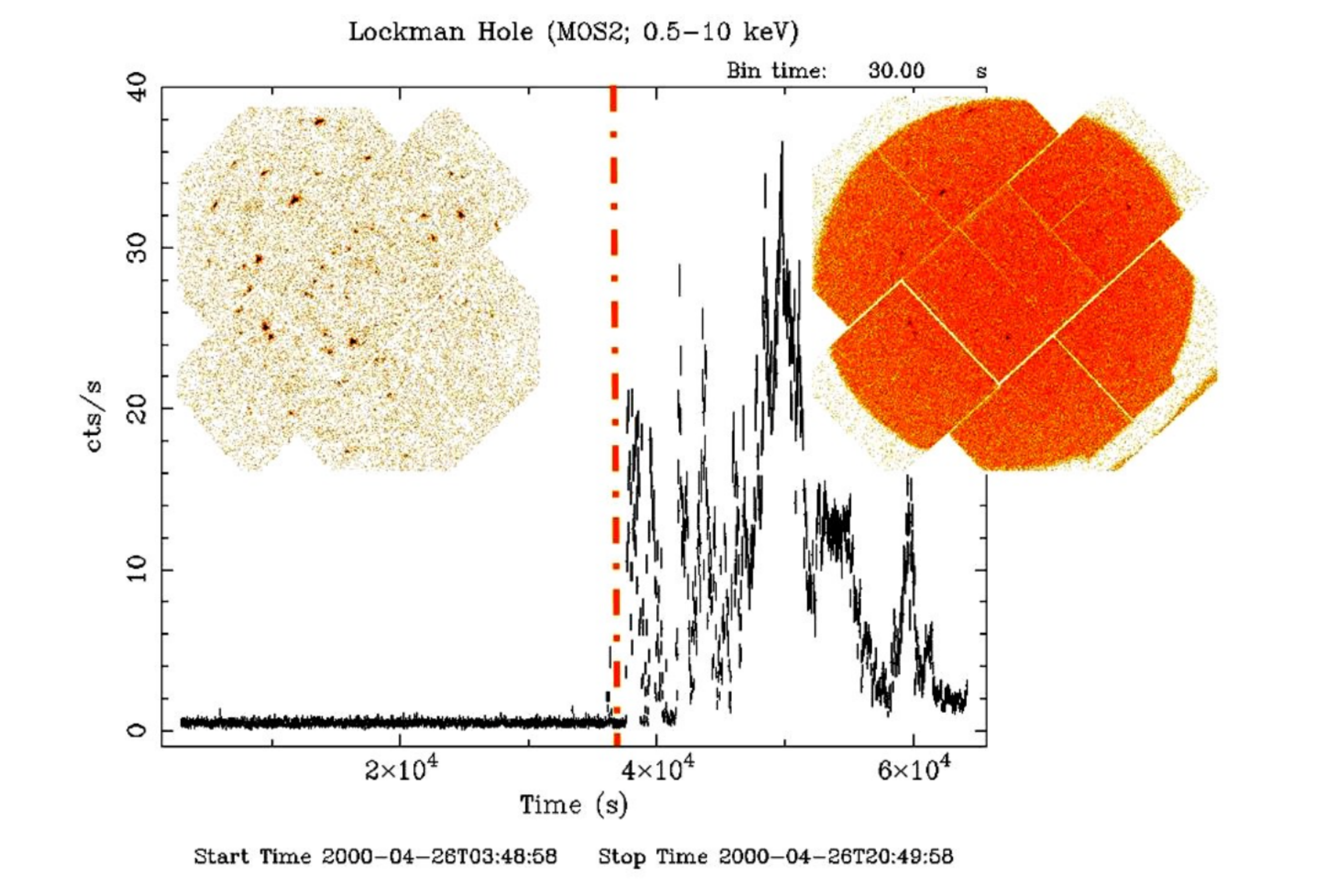}
	\caption{Example of XMM-Newton observation partly affected by soft protons. The flares are clearly visible in the second part of this light curve taken from MOS2, one of the detectors on board XMM. Their effect on the exposure quality can be evaluated comparing the image extracted from the first (left) and second half (right) of the observation.  Adapted from \citet{Lotti.ea:18}.   \label{fig:general}}
\end{figure}

XMM was launched  into  an orbit similar to Chandra, only  with apogee  in the southern hemisphere. To avoid radiation belts, the detectors of XMM are kept closed with a $\sim$1 mm thick aluminum shield below altitudes of about 40000 km. XMM highly eccentric elliptical orbit with an apogee of about 115000 km and a perigee of about 6000 km from Earth traverses the full range of magnetospheric environments, from the inner magnetosphere to the solar wind (SW) when the satellite is outside the bow shock. Along its orbit the satellite encounters enhanced intensities of SP.  These episodes are hereafter referred to ``soft proton flares''. They occur on extremely variable time scale, ranging from hundreds of seconds to several hours. The peak count rate can be more than three orders of magnitude higher than the quiescent one  \citep{Deluca04}. The extreme time variability is the fingerprint of this background component, so-called the SP component, which should not be confused with solar flares or solar energetic particles. A light curve can immediately show the time intervals affected by a high background count rate. Such intervals are usually not suitable for scientific analysis unless the X-ray source to be studied is extremely bright (see Figure \ref{fig:general}). They have to be rejected, discarding all of the time intervals having a count rate above a selected threshold. 

\begin{figure}[ht!]
	\includegraphics[width=0.9\linewidth]{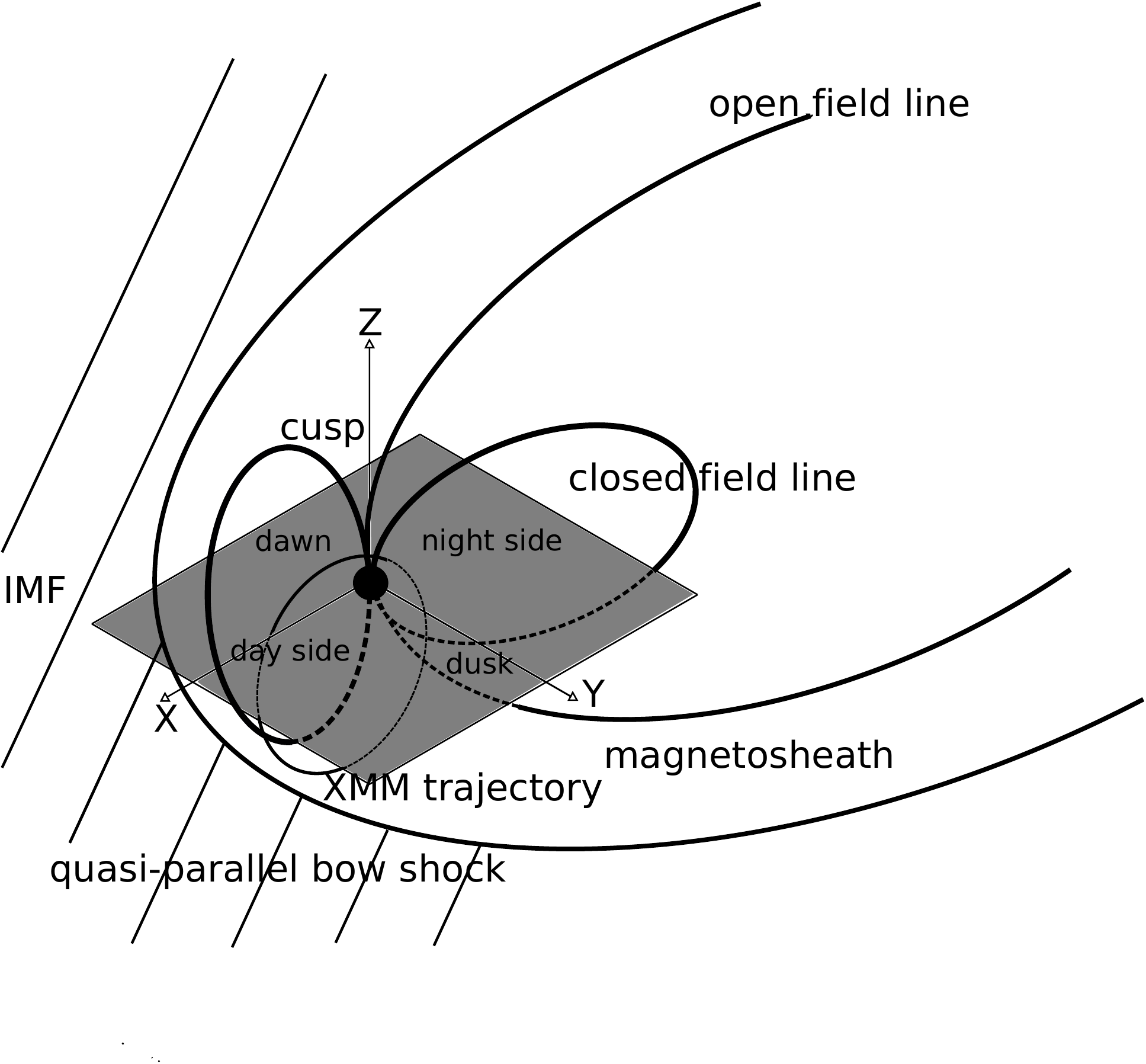}
	\caption{Sketch of the terrestrial magnetosphere, oblique lines in front of the magnetosphere represent Interplanetary Magnetic Field (IMF),  X, Y and Z denote directions of the Geocentric Sun Earth (GSE) coordinate system.  XMM apogee is found at $\sim$18 $R_{\mathrm{E}}$, where $R_{\mathrm{E}}$ is the Earth's radius. In the time period considered here, the XMM orbit has changed from highly elliptical to more circular and then back to highly elliptical.  \label{fig:sketch}}
\end{figure}

A preliminary analysis of the distribution of flares as a function of orbital position, distance from the Earth, and orbital phase with respect to the Sun has been done by \citet{Kuntz08}. The part of the orbit which seems the most susceptible to SP flare is in the inner part of the magnetosphere (near perigee), whereas greatest flare-free time occurs when the spacecraft is furthest from the Earth, either outside the bow-shock or deep within the magnetotail \citep{Ghizzardi:2017}. A development of that work based on XMM measurements from 2000 to 2010 for a 
total of 51 Ms of data concluded that the highest percentage of 
proton flares occurred when the spacecraft is on closed magnetic field lines \citep{Walsh14}, see sketch in Figure \ref{fig:sketch}. According to this study the SP affect $\sim$55\% of measurements and that can be as high as 66\% of measurements when XMM is located in low-latitude magnetospheric regions on closed magnetic field lines. Other studies as, e.g., \citet{Salvetti17} report a mean contamination rate $\sim$35\%.  A recent analysis based on about 100 Ms of data measured between 2000 and 2012 confirmed the general trend of a decreasing intensity with distance from Earth (shown by the mean count rate of the SP component). It also showed that the day-side magnetosphere with closed field lines is more contaminated by SP flares than regions on the night side on open field lines \citep{Ghizzardi:2017}. 

The performance of future X-ray focusing telescopes orbiting in the 
interplanetary space will suffer from SP induced background events. 
Of particular concern is European Space Agency's (ESA) next large class ATHENA \citep{Nandra.ea:13}
given that its large effective area (1.4 m$^2$ at 1 keV) makes the 
minimization of SP contamination a key challenge for the 
fulfillment of ATHENA's science objectives and it is explicitly recognized
in the background requirements of the mission. A possible shielding solution 
is placing an array of magnets (a magnetic diverter) between the optics and 
the focal plane, able to deflect charged particles away from the instruments 
field of view \citep[e.g.,][]{Fioretti.ea:18, Lotti.ea:18}. Also the initial
choice of an L2 orbit is being reconsidered due to the far superior knowledge
of the various proton components in L1 \citep{Fioretti.ea:18,Laurenza.ea:19}.   The first and second Sun-Earth Lagrange points (L1 and L2) are locations where the gravitational forces of the Sun and Earth cancel. Both L1 and L2 are located along the Sun-Earth line with L1 being 1.5 million kilometers Sunward of the Earth, while L2 is located at the same distance behind the Earth.

 In this paper we delineate which of the geometric, solar, SW, and geomagnetic parameters mostly control strong contamination in the XMM telescope using a Machine Learning (ML) approach. The eventual aim is to define the cause of the contamination. ML approach has been successfully used to predict plasma environments in the terrestrial magnetosphere such as electron density in the plasmasphere \citep{Zhelavskaya17} and the inner magnetosphere \citep{Chu17}, the electron intensity in the radiation belts \citep{Smirnov20} as well as in the solar wind \citep{Roberts20}. The advantage of this approach is that it allows complex non-linear relationships to be analyzed in large datasets \citep{geron19}. Our task is to predict target numeric values, namely the count rate of the SP contamination, given a set of features, such as, location of the satellite, solar, SW and geomagnetic parameters, called predictors. We treat this problem as a regression (see, e.g., \citet{Camporeale19} for details).  To train the algorithm, one feeds it with many examples of events that include both their predictors and their desired solutions (count rates of the contamination in our case). Such ML approach is called supervised learning; the training set given to the algorithm includes the desired solution. Some of the most important supervised learning algorithms are linear regression, Support Vector Machines (SVMs), Decision Trees and Random Forests (RF), Neural networks, Gradient Descent and Gradient Boosting (GB).

To predict the contamination, we first explored the relation with the single parameters
to help  select  the best predictors for an ML model.
With this choice we test a row of supervised ML algorithms and eventually
derive a model which utilizes an ensemble of predictors based on the Extra Trees
Regressor algorithm. Using this ML algorithm we evaluated importance of non-linear relationships.

The ML approach may help in searching for similar patterns between the
XMM contamination and SP intensities measured by Cluster, thus constraining the source of SP. ESA's mission Cluster  which is a suite of four satellites  \citep{Escoubet97} orbits the Earth on polar trajectories similar to XMM. However, there are no physical conjunctions between these
two satellites that would allow direct insights on what exactly produces
contamination. Therefore, in the future one approach will be the identification of possible magnetic field conjunctions (observations at similar magnetic field topologies) and comparing observations from both missions. Another approach  will be  to delineate which geometric, solar,
SW, and geomagnetic parameters are most related to dynamics of SP at
different energies observed by Cluster to compare with those parameters associated with the XMM contamination.  In the future, we  will derive a ML predicting model for the SP measured by Cluster and apply it to XMM trajectories to disentangle at which energies SP are the best correlated with the contamination. By this we will determine the energy of SP that contaminates the detector the most. 

This work has been inspired by the
interdisciplinary collaboration between the astrophysicists and specialists in
the magnetospheric physics supported by the International Space Science
Institute in Bern, Switzerland\footnote{\url{https://www.issibern.ch/teams/softprotonmagxray/}}.

\section{Contamination SP count rates and their predictors: simple relations} \label{sec:data}

In this Section we give details about SP contamination count rates and their predictors. We plot their relations and analyse cross-correlations in order to get better insights into physical processes possibly responsible for the contamination and to have better preselection of the predictors for the ML model.

\subsection{Contamination count rates}

The description of the XMM dataset and the analysis performed have been 
reported in more details in \citet{Marelli.ea:17} and \citet{Salvetti17}.
Here we give a brief and concise summary for the purpose of this paper.
The work exploited here have been produced in the framework of AREMBES (ATHENA Radiation 
Environment Models and X-Ray Background Effects Simulators) which is an ESA project
aimed at characterizing the effects of focused and non-focused particles on 
ATHENA detectors: both in terms of contributions to their instrumental 
background and as source of radiation damage\footnote{http://space-env.esa.int/index.php/news-reader/items/AREMBES.html}.
XMM-Newton is a test-bed of the various background components which will be 
relevant for the ATHENA mission. To this aim we  used  the XMM-Newton 
public data set   which  was  available when we started  our analysis  to produce the most
clean data set ever used to characterize the XMM-Newton particle-induced 
background, taking as input the preliminary results of the FP7 European 
project EXTraS (Exploring the X-ray Transient and variable Sky\footnote{http://www.extras-fp7.eu/} \citep{De-Luca.ea:15}).  The results from the Data Release 4 of the 3XMM catalogue were required to evaluate the contamination from celestial sources. 

The main XMM instrument is the European Photon Imaging Camera (EPIC), 
consisting of two  Metal-Oxide-Silicon (MOS) detectors \citep{Turner01} and a pn camera \citep{Struder.ea:01} which  operate  in the 0.2--12 keV energy range. The EPIC 
background can be separated into particle, photon and electronic noise 
components (see \citet{Carter07} and \citet{Gastaldello.ea:17} for a 
detailed description). Aiming to characterize the SP component which 
is focused by the X-ray telescopes, the key feature exploited is the ability 
to define in the MOS detectors two detector areas: the 
in-Field-Of-View (inFOV) one, exposed to focused X-ray photons and SP, and the out-Field-Of-View (outFOV) one, not exposed to sky photons 
nor SP. The other main component of the particle background, 
secondary electrons generated by Galactic Cosmic Rays affects in the same way
both inFOV and outFOV areas of the MOS detectors. The choice of the energy band
in the analysis (7--9.4 and 11--12 keV) minimizes to a negligible contribution 
the sky photon component. We focused on MOS2 because
we can exploit the full detector area (MOS1 suffered loss of 2 of its 7 CCDs during the lifetime of the mission).

We can then use the inFOV subtracted by outFOV diagnostic to fully characterize
the inFOV excess particle background employing the outFOV region as a 
calibrator to minimize any contamination. After standard data preparation and 
reduction, all the single observations were merged in a final global dataset used in this work
with 500 s time bins, where the count rate is the difference between the inFOV 
and outFOV count rate. The work
done in the AREMBES project showed two distinct components in the differential
distribution of the inFOV-outFOV count-rates, one associated to the flares
of SP and the other to a low intensity component, possibly related
to Compton interactions of hard X-ray photons. This fundamental distinction
is supported by the comparative analysis of data collected with different 
filters and a spectral analysis \citep{Salvetti17}.

 We investigate the dynamics of the SP count rates between 0.04 and 200  counts/s. We slightly revise the lowest threshold of 0.1  counts/s used
in \citet{Ghizzardi:2017}.  A threshold of 0.1 was chosen in \citet{Ghizzardi:2017} to be in a regime totally dominated by the SP contribution. However, the regime between 0.04--0.10 still provides a significant contribution with respect to the other major component of the XMM background which is the Galactic-Cosmic ray induced background (which ranges in the same units from 0.1 to 0.4, see  Figure 4 left panel of \citet{Salvetti17}).
We select the observations with radial distance above $6 R_{\mathrm{E}}$. From January, 2 2001 to  August 30, 2012, 707 330 minutes of data matched these criteria. We also applied the base 10 logarithm to the SP count rates because the data variation is in the range of several orders of magnitude. The distribution of the number of samples  for the predictors and the count rates  (on the vertical axis)  with  a given value range (on the horizontal axis) is shown in Figure  \ref{fig:hist} in Appendix.

\subsection{Predictors related to location in space}

\begin{figure}[ht!]
\plotone{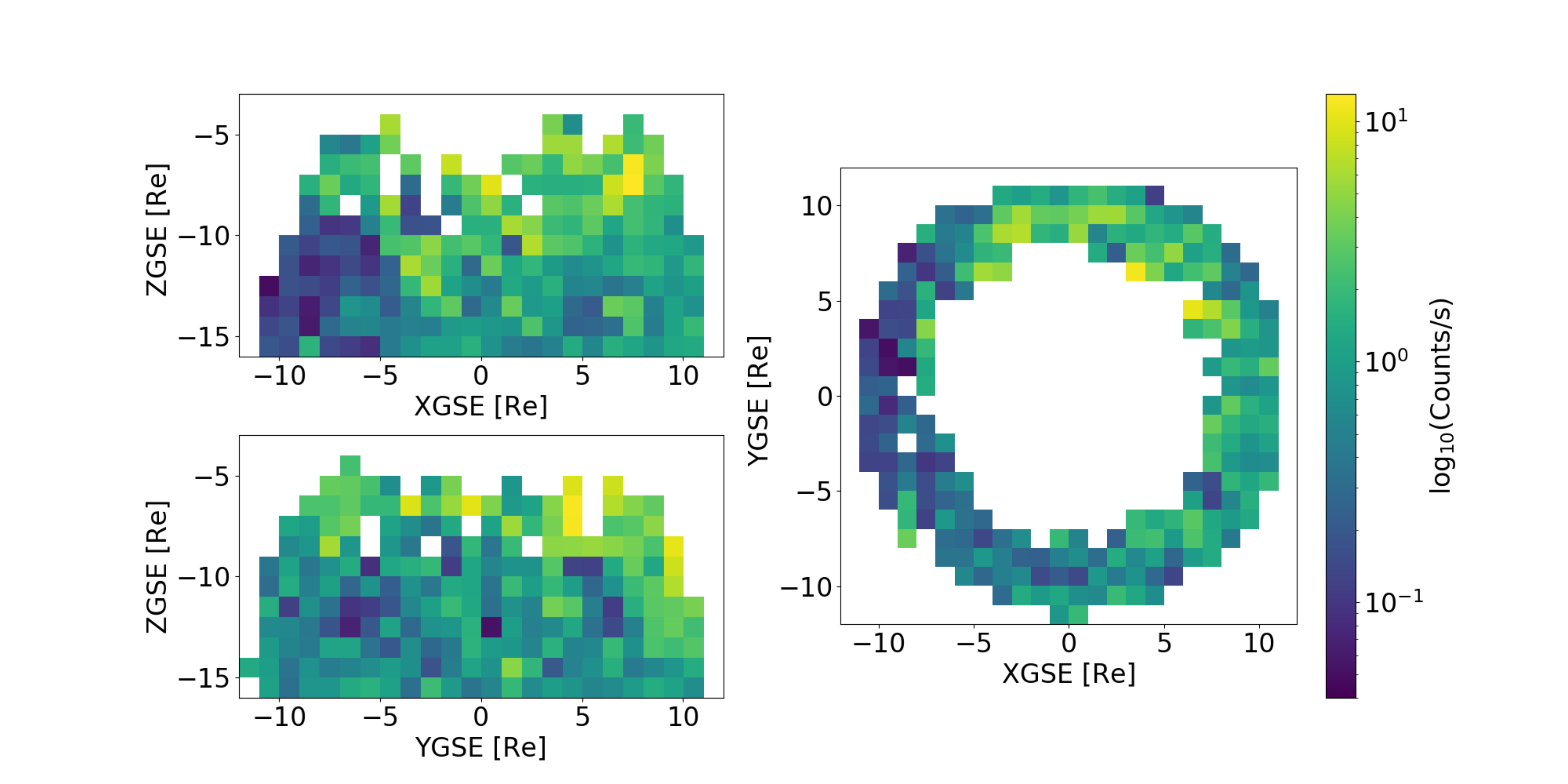}
\caption{Distribution of the SP count rates in the range between 0.04 to 200 counts/s in the GSE coordinate system. Number of the SP count rates per bin is larger than 2.  \label{fig:distr}}
\end{figure}

Each count rate was associated with location in GSE coordinate system represented by parameters $X, Y, Z$ (see sketch in Figure \ref{fig:sketch}) and the radial distance from the Earth, parameter $rdist$. Throughout the paper distances are given in $R_{\mathrm{E}}$ units. The distribution of the SP count rates in the GSE system  is  shown in  Figure  \ref{fig:distr}.  The figure  shows that the dayside  (positive XGSE)  is more affected by the contamination. A duskward asymmetry is observed, with stronger contamination towards flanks on the dusk side, higher count rates at approximately YGSE=8 R$_E$ and XGSE 6--12 R$_E$  (see sketch in Figure \ref{fig:sketch} that indicates location of the dusk/dawn and day/night sides). Figure \ref{fig:distr} illustrates a decrease of SP contamination at larger distances from the Earth in $Z$ direction.

\begin{figure}[ht!]
	\includegraphics[width=0.8\linewidth]{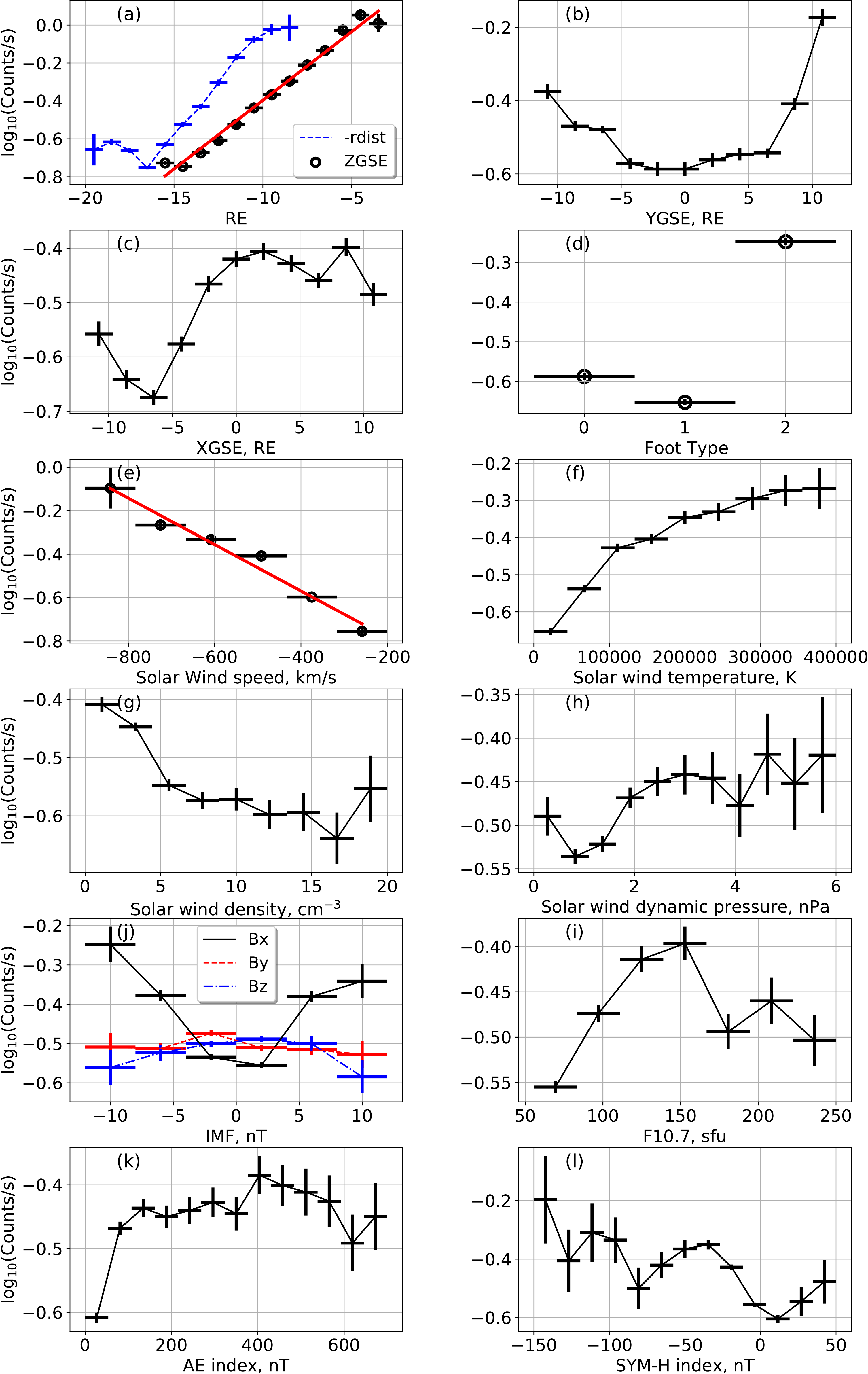}
            \caption{Relations of  mean  XMM count rates and (a)-(c) ZGSE together with linear regression shown by the red line and negative radial distance, YGSE and XGSE, respectively; (d) Foot Type; (e) the solar wind radial velocity and its linear regression shown by the red line; (f)-(h) solar wind temperature, density and dynamic pressure, respectively; (j) IMF components in GSE; (i) F10.7 parameter; (k) AE index and (e) SYM-H index. Vertical lines represent standard in statistics confidence intervals at 95\% confidence level.  Horizontal lines represent the half width of the bin for which the corresponding values were calculated. The data points are connected by thin lines to guide the eye. \label{fig:binned}}
\end{figure}

In Figure \ref{fig:binned} we plot count rates versus individual predictors.
One can see that the logarithm of the SP count rates almost linearly decreases with $Z$, see Figure (\ref{fig:binned}, a). This dependence is the strongest compared to the other parameters, considering the span of the count rate values.  The linear regression derived from this dependence (shown by the red line in Figure \ref{fig:binned}, a) is
$$\log_{10}(\text{SP Count rates})=0.328 +0.0725\cdot Z.$$
This relation indicates an exponential dependence of the SP count rates on $Z$. For this linear regression the Pearson correlation, $r,$  is 0.99, probability value, $p,$ is 4$\cdot 10^{-11}$ and standard error of the estimated gradient is 3$\cdot 10^{-3}$. 

The change of the logarithm of count rates with $Y$ is non-linear and significantly weaker than those on $Z$, see Figure (\ref{fig:binned}, b). The expected from Figure  \ref{fig:distr} duskward asymmetry is clearly observed. A slightly less strong, non-linear, dependence of logarithm of count rates is seen with respect to $X$  (Figure \ref{fig:binned}, c), with a higher level of contamination along the dayside, as also seen from Figure  \ref{fig:distr}. 

 Previous studies \citep[e.g.][]{Walsh14} have demonstrated that the XMM SP contamination count rates depend on the type of the connection of the magnetic field line to the Earth. Therefore, we have added a parameter called \textit{Foot Type}: closed magnetic field lines with both ends at the Earth (\textit{Foot Type} = 2), open magnetic field lines with one end at the Earth  and the other end connected to the Interplanetary Magnetic Field (IMF)  (\textit{Foot Type} = 1) and magnetic field lines are not connected to the Earth, namely IMF (\textit{Foot Type} = 0), see sketch in Figure \ref{fig:sketch}.  The parameter \textit{Foot Type} also describes the location of the contamination with respect to the Earth's magnetosphere.  This parameter was calculated using Tsyganenko 96 model \citep{Tsy95}.  There are later versions of the Tsyganenko model that may be more appropriate in periods of high solar wind dynamic pressure. However, for practical reasons we use only one model. In Figure (\ref{fig:binned}, d) we can see there is significantly higher count rates on the closed field lines (Foot Type = 2), than either on open field lines (Foot Type = 1) or on IMF field lines (Foot Type = 0). Additionally, the count rates in the IMF (Foot Type = 0) are also significantly higher than on the open field lines (Foot Type = 1).

 \subsection{Predictors related to the solar, solar wind and geomagnetic activity}

  The XMM count rates were combined with simultaneous observations of the solar, SW and geomagnetic parameters taken from the OMNI data base\footnote{\url{https://omniweb.sci.gsfc.nasa.gov}}, see also \citet{Papitashvili05}. The SW observations  are taken  from the OMNI data set.  They are propagated to the Earth's bow shock.  The SW is characterized by the proton density, \textit{NpSW} in $cm^{-3}$  (see Figure (\ref{fig:binned}, g)); components of the speed in the GSE coordinates, \textit{VxSW\_GSE}, \textit{VySW\_GSE} and \textit{VzSW\_GSE} in km s$^{-1}$ (see Figure (\ref{fig:binned}, e) for the former component); the temperature, \textit{Temp}, in $K$ (see Figure (\ref{fig:binned}, f)); the dynamic pressure, \textit{Pdyn} in nPa, which is calculated as \textit{NpSW*VSW}$^2\times1.67\cdot 10^6$ (see Figure (\ref{fig:binned}, h)); components of the IMF in the GSE coordinates, \textit{BimfxGSE},  \textit{BimfyGSE} and  \textit{BimfzGSE} in nT  (see Figure (\ref{fig:binned}, j)) and Clock Angle (\textit{CA}) calculated as  $\arctan$(\textit{BimfyGSE}/ \textit{BimfzGSE}).  To consider the influence of solar irradiation we included the F10.7 index which measures the radio flux at 10.7 cm (2.8 GHz)  \citep{Tapping13}. This parameter correlates well with the sunspot number and other indicators of solar and UV solar irradiance and can be measured reliably under any terrestrial  weather condition (unlike many other solar indices). It is denoted by \textit{F107} and measured in solar flux units (sfu)  (see Figure (\ref{fig:binned}, i)).  The parameters of geomagnetic activity such as Auroral Electrojet (AE) index, denoted as \textit{AE\_index}, in nT, characterizing the magnetic field disturbance  in the auroral  region of the northern hemisphere   and SYM-H index, denoted as \textit{SYM-H} and measured in nT, characterizing the disturbance of the geomagnetic field at the equatorial regions, are considered  \citep{AEindex}.  
  
  In Figure \ref{fig:binned} we plot parameters for most prominent relations with the  SP  count rates. We acknowledge that the SW and geomagnetic properties are correlated with one another. Thus, we try to determine here which dominates.
 
The logarithm of count rates increases almost linearly with absolute value of the SW speed, see Figure (\ref{fig:binned}, e). The linear regression derived from this dependence (shown by the red line in Figure \ref{fig:binned}, e) is
$$\log_{10}(\text{SP Count rates})= -0.997-10^{-3}\cdot V_x.$$
This relation indicates an exponential dependence of the SP count rates on $V_x$. The $r$ is -0.99, $p$ value is 2.5$\cdot 10^{-4}$ and standard error of the estimated gradient is 8.7$\cdot 10^{-5}$. 

The count rates clearly increase with the SW temperature, see Figure (\ref{fig:binned}, f). The count rates non-linearly decrease with the SW density, see Figure (\ref{fig:binned}, g). The SP count rate relation with the SW pressure is non-linear, see Figure (\ref{fig:binned}, h). For SW pressures higher than 6 nPa, the confidence intervals cover the entire value range of count rates. These values are discarded because they are statistically insignificant. The relation of count rates with the SW pressure is less important than the one with the SW speed because the count rates anti-correlate with the SW density,  see Figure (\ref{fig:binned}, g). SW speed is often anti-correlated with SW density \citep{Richardson18}, therefore reducing the significance of the SP count rates relative to the SW pressure. 

Of the IMF components, the $B_x$ component, shows the strongest relation with the logarithm of count rates, see Figure (\ref{fig:binned}, j). The XMM count rates increase with the absolute value of the IMF $B_x$ component.  The IMF $B_y$ component does not show significant change. It is rather unexpected to observe a significant decrease of the SP count rates for absolute values of IMF $B_z>$8 nT. The strong values of IMF  $|B_z|>$8  nT are likely associated with   geoeffective interplanetary phenomena such as Coronal Mass Ejections (CMEs) and Corotating Interaction Regions (CIRs) \citep{Gonzalez99,Mcpherron06,Li18},  and intuitively one would expect an increase in the count rates. This will be discussed in Section \ref{sec:discussion}.

The logarithm of count rates first increases linearly with the F10.7 index up to 150 sfu and then significantly drops at higher values, see Figure (\ref{fig:binned}, i), indicating a non-trivial influence of this parameter on level of contamination. 

The dependence of the contamination on the substorm activity, indicated by AE index, is weaker than for the SW speed, see Figure (\ref{fig:binned}, k). The SP count rates significantly grow with AE index at least up to 100 nT. In general, AE index, namely strong magnetic field disturbance in the northern  high-latitude region, does not show a significant relation with count rates at values $>$100 nT.

The dependence of the count rates on the SYM-H index is also non-trivial. The SP count rates increase for decreasing values of the SYM-H index from 0 up to approximately -50 nT, see Figure (\ref{fig:binned}, l). At lower SYM-H index values dependence becomes non-linear with large error bars.  An increasingly negative SYM-H index means that the ring current is stronger at equatorial latitudes.

\subsection{Cross-correlations between contamination counts and predictors}

\begin{figure}[ht!]
 	\plotone{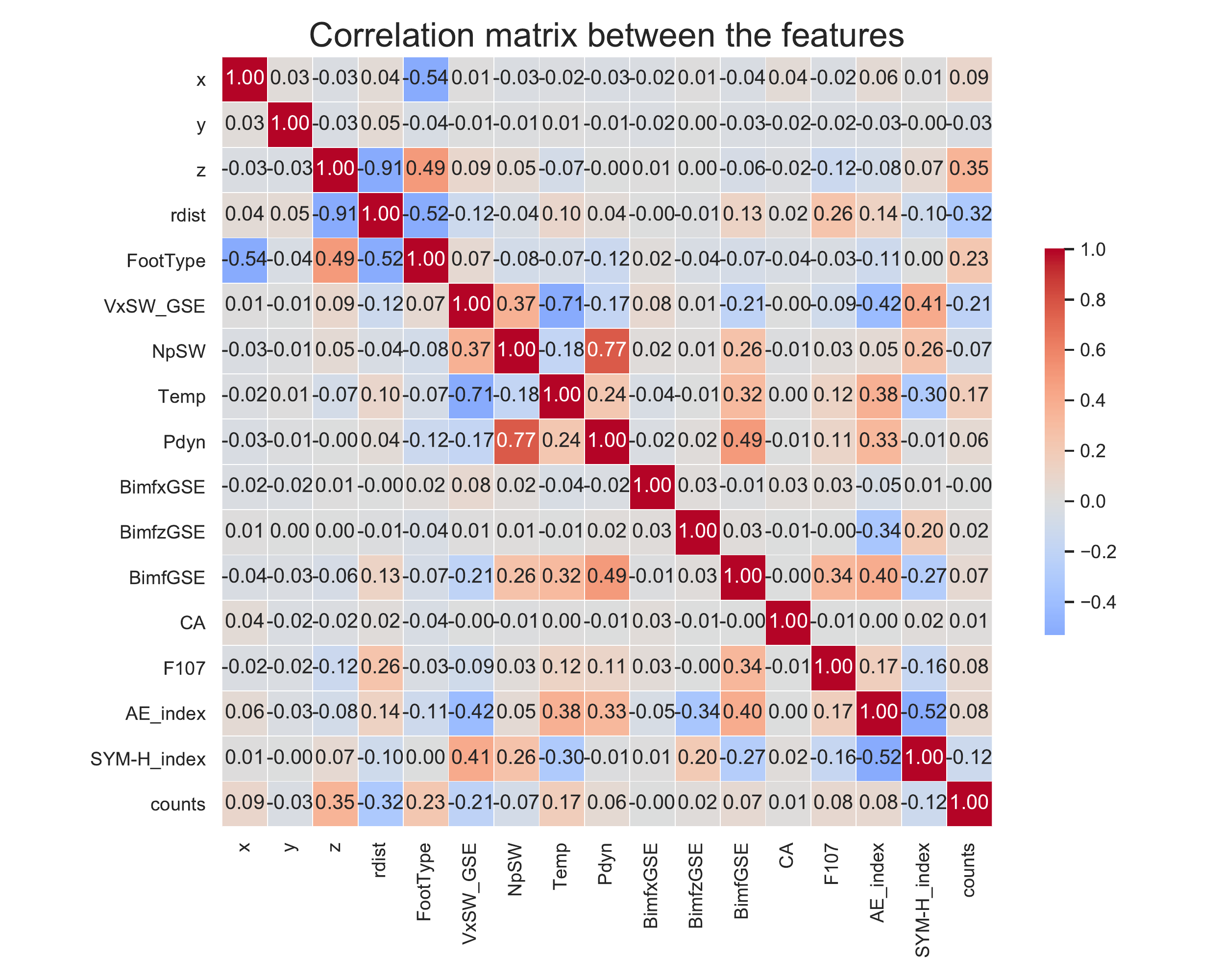}
 	\caption{Correlation matrix between parameters. Here we used the Pearson correlation. The correlations are rounded to the second decimal for better vizualization.   \label{fig:corr}}
 \end{figure}

In Figure \ref{fig:corr} we show the correlation coefficient, $r,$ between parameters possibly related the level of SP contamination. The values of Pearson correlation vary between -1 and 1, with values close to -1/1 meaning perfect linear anticorrelation/correlation and values close to 0 meaning no linear correlation. We dropped the  \textit{VySW\_GSE} and \textit{VzSW\_GSE} components from this plot for the sake of better presentation as they show very low correlation with SP counts and small influence on reproducing the counts in the model. These velocity components are small compared to the \textit{VxSW\_GSE}. We also checked correlation and influence of the total SW speed on the reproducibility of the SP count rates, however, it shows a very similar behavior to \textit{VxSW\_GSE}. In order to avoid redundant parameters we do not include this variable. Additionally, we dropped  \textit{BimfyGSE,} from Figure  \ref{fig:corr} due to the low Pearson correlation and no obvious relationship  with SP counts in Figure \ref{fig:binned}. The correlations help us to exclude parameters that are strongly correlated with one another that can overload the model. However, one should be also careful with the interpretation of the Pearson correlation coefficient as this indicates only linear relationships. 

  The results of the cross-correlation analysis  follow the strongest relations with count rate,  such as variation with ZGSE direction ($r$=0.35), the radial distance ($r$=0.32), Foot Type ($r$=0.23), \textit{VxSW\_GSE} ($r$=0.21), SW temperature ($r$=0.17) and  SYM-H index ($r$=0.12). Here we chose predictors with correlation larger than 0.1. However, one can also note well defined non-linear relations of count rates with $X$, $Y$ and \textit{BimfxGSE}, in Figure \ref{fig:binned} that got low scores in Pearson correlation. 

On the basis of correlations and dependencies in Figure \ref{fig:binned} we select the following predictors for the ML model: $X$, $Y$, $Z$, \textit{rdist}, \textit{Foot Type}, \textit{VxSW\_GSE}, \textit{Pdyn}, \textit{BimfxGSE}, \textit{F107}, \textit{AE\_index} and \textit{SYM-H}. The  \textit{BimfyGSE} and \textit{BimfzGSE} are dropped because they do not show much variation with the counts. The  \textit{rdist}/\textit{NpSW}/\textit{Temp}  are dropped because they correlate strongly with  $Z$/\textit{Pdyn}/\textit{VxSW\_GSE}  and do not  significantly  improve the model. 

\section{ML model for SP contamination} \label{sec:model}

The relation between the SP count rates and the row of different predictors listed above is complex, see Figure \ref{fig:binned}. It is, therefore, often a group of predictors or their ensemble that gives better predictions than the best individual predictor \citep{geron19}. 

From supervised ML regressions we have tried Stochastic Gradient Descent Regressor (\verb|SGDRegressor|), Gradient Boosting for Regression (\verb|GradientBoostingRegressor|),  Random Forest Regressor (\verb|RandomForestRegressor|), Extra Trees Regressor (\verb|ExtraTreesRegressor|) and Multi-layer Perceptron  Regressor  (\verb|MLPRegressor|) methods implemented in Scikit-Learn  \citep{scikit-learn}.  These methods show comparable or slightly worse performance, see Table \ref{tab:messier}. To evaluate performance we use Spearman correlation, $\rho,$ between results of the model on training/test datasets and observations that are listed as Train Spearman/Test Spearman in Table \ref{tab:messier}, respectively. The values of Spearman correlation vary between -1 and 1, with values close to -1/1 meaning perfect linear anticorrelation/correlation and values close to 0 meaning no linear correlation. Although, Gradient Boosting Regressor has shown slightly better predicting performance and is less inclined to overfitting (scores for training of the model and evaluation are similar, see discussion below), we have decided to use Extra-Trees Regressor because it gives more consistent results between estimators (see below) and it is computationally more efficient. This method works well on noisy data \citep{Geurts06}.

Extra-Trees Regressor is an ensemble learning method that constructs multiple decision trees during training and outputs a mean prediction of the individual trees. This algorithm builds an ensemble of regression trees
according to the classical top-down procedure. Two main differences with other tree based ensemble methods are that it splits nodes using random thresholds for each feature rather than searching for the best possible thresholds and
that it utilizes the whole learning sample (compared to a bootstrap replica, namely resampling a dataset with replacement) to grow the trees \citep{Geurts06, geron19}.  We use Extra-Trees Regressor implemented in Scikit-Learn function \verb|ExtraTreesRegressor| version 0.22.1.

\begin{deluxetable*}{lccc}
	\tablenum{1}
	\tablecaption{Performance of different models with default input values. \label{tab:messier}}
	\tablewidth{0pt}
	\tablehead{
		\colhead{Regressor} & \colhead{Train Spearman} &  \colhead{Test Spearman}}
	\startdata
	Extra Trees&    1.000&   0.441\\
	Random Forest&     0.947 &   0.402\\
	Gradient Boosting& 0.565&    0.452\\
	Multi-layer Perceptron   &   0.605&    0.439\\
	Stochastic Gradient Descent&    0.447&    0.408\\
	\enddata
\end{deluxetable*}
 
 \subsection{Training the model} \label{sec:modeltrain}

 The XMM SP count rates data set consists of data from January 2, 2001 to August 30, 2012. We took the data for training of the model and its validation from January 2, 2001 to December 31, 2010. The rest of the data from January 1, 2011 to August 30, 2012 is used only for the testing of the model. The ratio between amount of data for training/validation and testing is about 10:1.8. This is standard partitioning in ML \citep{geron19}.
 
The parameters we want to correlate with the SP count rates all have different  ranges, therefore we  decided  to scale the data. We tried Scikit-Learn functions such as  \verb|StandardScaler|,  \verb|RobustScaler|, \verb|MinMaxScaler| and \verb|QuantileTransformer|.  The latter scaler gave the best performance and is used in our model.

We trained the model using $K$-Folds cross-validation (function \verb|model\_selection.KFold|) with number of splits equal 5. This method  randomly divides  the data set into $K$ subsets of approximately the same size called folds, in our case $K=5.$ Then we train and evaluate the Extra-Trees Regressor model 5 times by choosing a different fold for evaluation every time and training on the other 4 folds. This results in 5 arrays of evaluation scores. The advantage of training the model several times is that we can derive average performance of the model for the train/validation data set, considering that in our case  we observe a wide dynamic range in SP count rates. Another advantage is that one can estimate the precision of the model by deriving, e.g., its standard deviation.

 We use 1 to 200 trees with depths in the range from 1 to 20. Other parameters in the \verb|ExtraTreesRegressor| were set as default.  To evaluate the performance of the training and validation during cross-validation for different parameters we use four different assessment metrics: Spearman correlation  ($\rho$),  Mean Square Error (MSE), Mean Absolute Error (MAE) and coefficient of determination  ($R^2$).  The values of MSE and MAE tend to zero in case of perfect agreement between the model and observations. $R^2$ indicates which fraction of data variability the model can explain, in the perfect case it is equal to 1.

\begin{figure}[ht!]
	\plotone{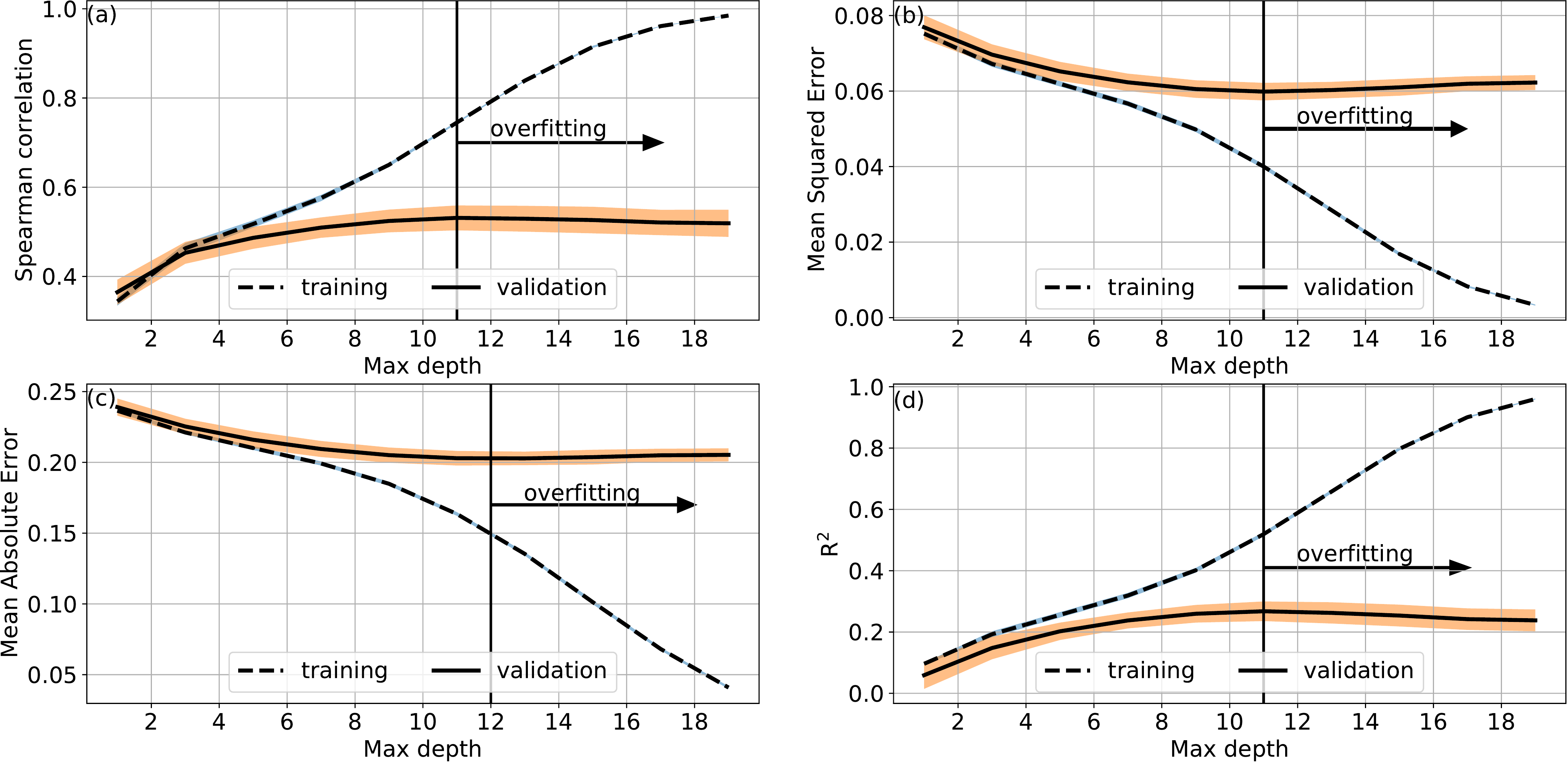}
	\caption{Performance of the model for metrics $\rho$, MSE, MAE and $R^2$  ((a) to (d))  versus depth of the trees for averaged training (solid line) and validation (dashed line) data sets. The number of estimators is equal to 130. The blue and orange color indicate standard deviation for 5 cross-validation evaluation scores. \label{fig:performance_error}}
\end{figure}

To select best parameters of the estimator we used optimization by cross-validated grid-search over a parameter grid, \verb|GridSearchCV|. This was done for four different metrics $\rho$, MSE, MAE and $R^2$. The performance of the model for the training/validation data sets is consistent between different metrics. The highest performance is observed for $\simeq$130  trees  and   depth  of 12 for MAE and 11 for MSE, $\rho$ and $R^2$. We plot the performance metrics of the model versus depth of the trees for the training and validation data sets for 130  trees  in Figure  \ref{fig:performance_error}. In the figure by vertical line we indicate the optimal depth of the trees when a metric shows the minimum of validation error. For the depths of the trees with higher values the model starts to overfit the data \citep{Prechelt98}. Namely, the discrepancy in performance between the training and validation data sets becomes larger \cite[e.g.][]{Ghojogh19}. In the ideal case the gap between training and validation errors should be small \citep{Goodfellow17}. For the model we select the depth of the trees equal to 11, the value at which the approximate minimum of validation error is observed . At this value the gap between training and validation errors is not too large yet.

The model is stable to outliers. We tried to limit the ranges of parameters, however this did not improve the performance of the model significantly. We, therefore, do not limit the ranges of predictors. 

\begin{figure}[ht!]
\includegraphics[width=0.48\linewidth]{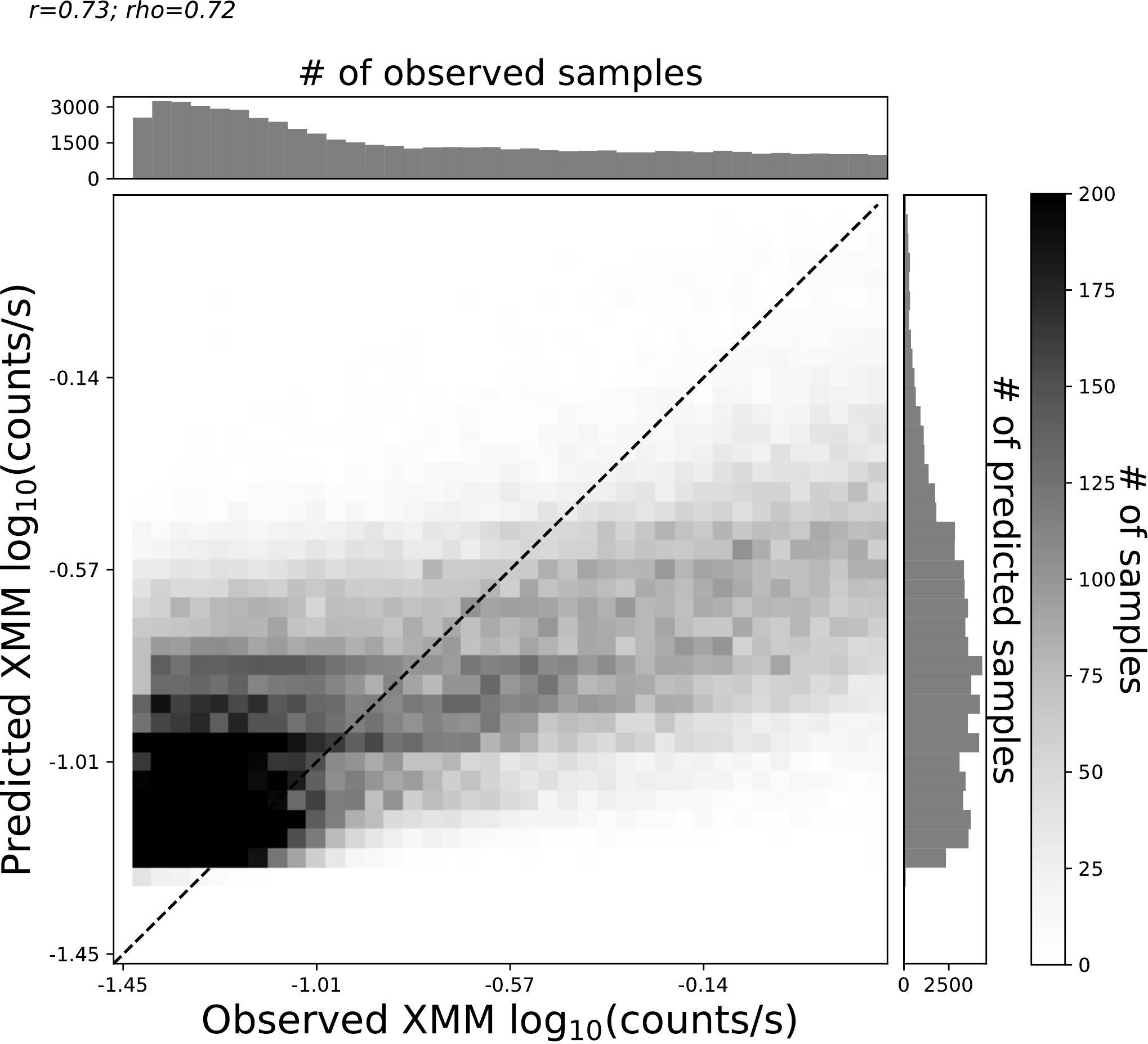}
\hspace{4mm}
\includegraphics[width=0.48\linewidth]{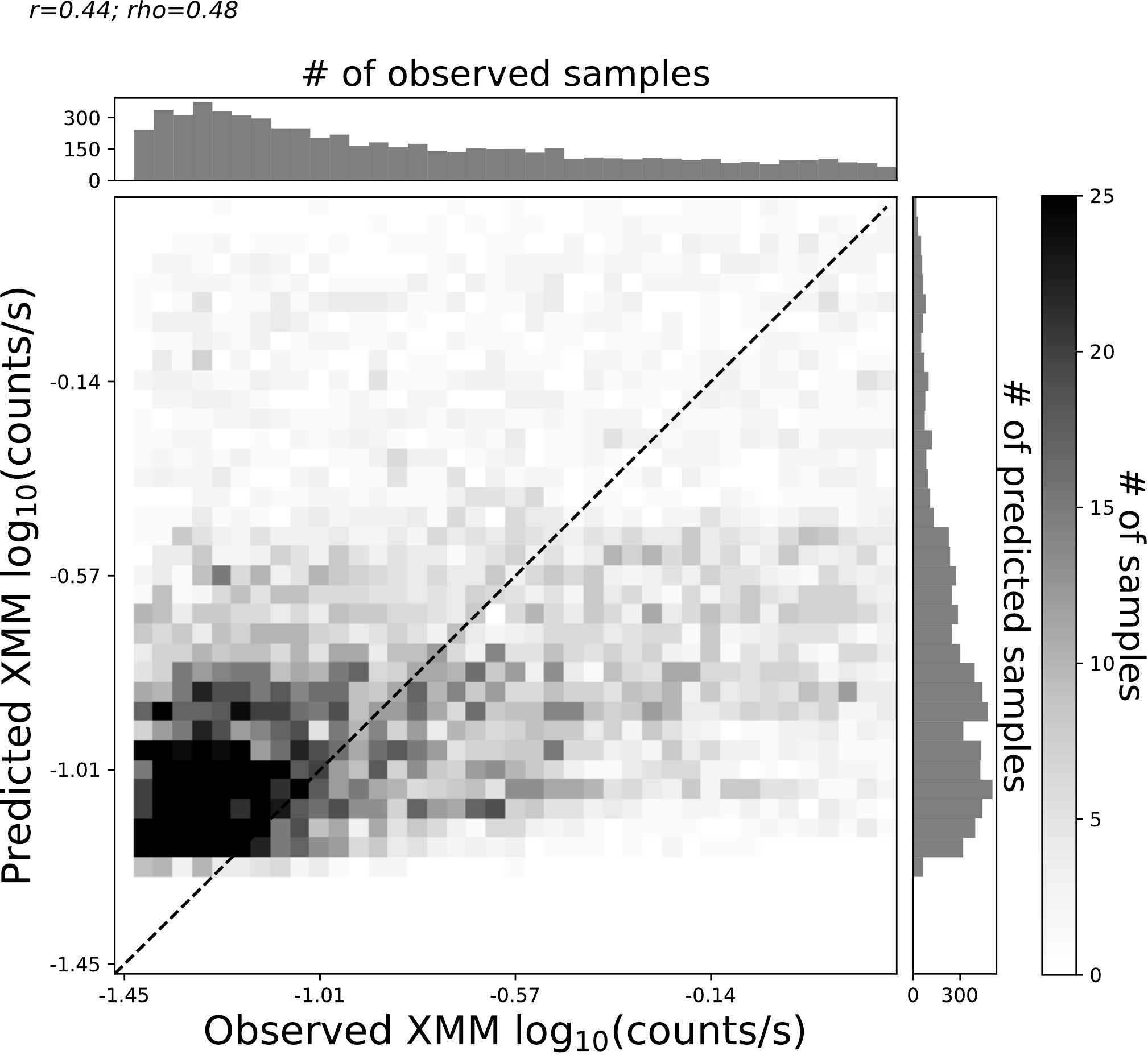}
\caption{Observed count rates from the trained (left) and test (right) data set versus those predicted by the model. The color represents number of samples. \label{fig:trained}}
\end{figure}

The distribution of the observed count rates versus predicted by the model based on trained data set is shown in Figure \ref{fig:trained} (left)  and will be discussed in Section \ref{sec:results}.  The performance of the trained model evaluated by different estimators is listed in Table \ref{tab:performance}.  

\begin{deluxetable*}{lccccc}
\tablenum{2}
\tablecaption{Performance of the ML and linear models for trained/validation and test data sets. \label{tab:performance}}
\tablewidth{0pt}
\tablehead{
\colhead{Data set} &\colhead{$r$} & \colhead{$\rho$} &  \colhead{MSE} &  \colhead{MAE}&  \colhead{$R^2$}}
\startdata
ML Train& 0.72& 0.72&0.04&0.17& 0.49\\
ML Test&0.47&0.48&0.06&0.2& 0.18 \\
linear Z Train data set&0.35&0.32&0.45&0.57&0.12\\
linear Z Test data set&0.28&0.26&0.41&0.54&0.03\\
linear $V_x$ Train data set&0.19&0.20&0.42&0.55&0.01\\
linear $V_x$ Test data set&0.18&0.16&0.42&0.55&0.02\\
\enddata
\end{deluxetable*}

\subsection{Predictor importances}

This method provides an opportunity to assess the relative importance of a feature with respect to the predictability of the target variable. This corresponds to the relative rank (tree depth) of a predictor used as a decision node in a tree. Features at the top of the tree affect the final prediction decision of a larger number of input samples. The relative importance of the predictors is the expected fraction of the samples they affect. One can average the estimates of predictive ability over several randomized trees. This will reduce the variance of such an estimate and is called the mean decrease in impurity. In Scikit-Learn, the relative importance is combined with the mean decrease in impurity forming a normalized estimate of the predictive power of that feature \citep{scikit-learn, Louppe14}.

\begin{figure}[ht!]
	\plotone{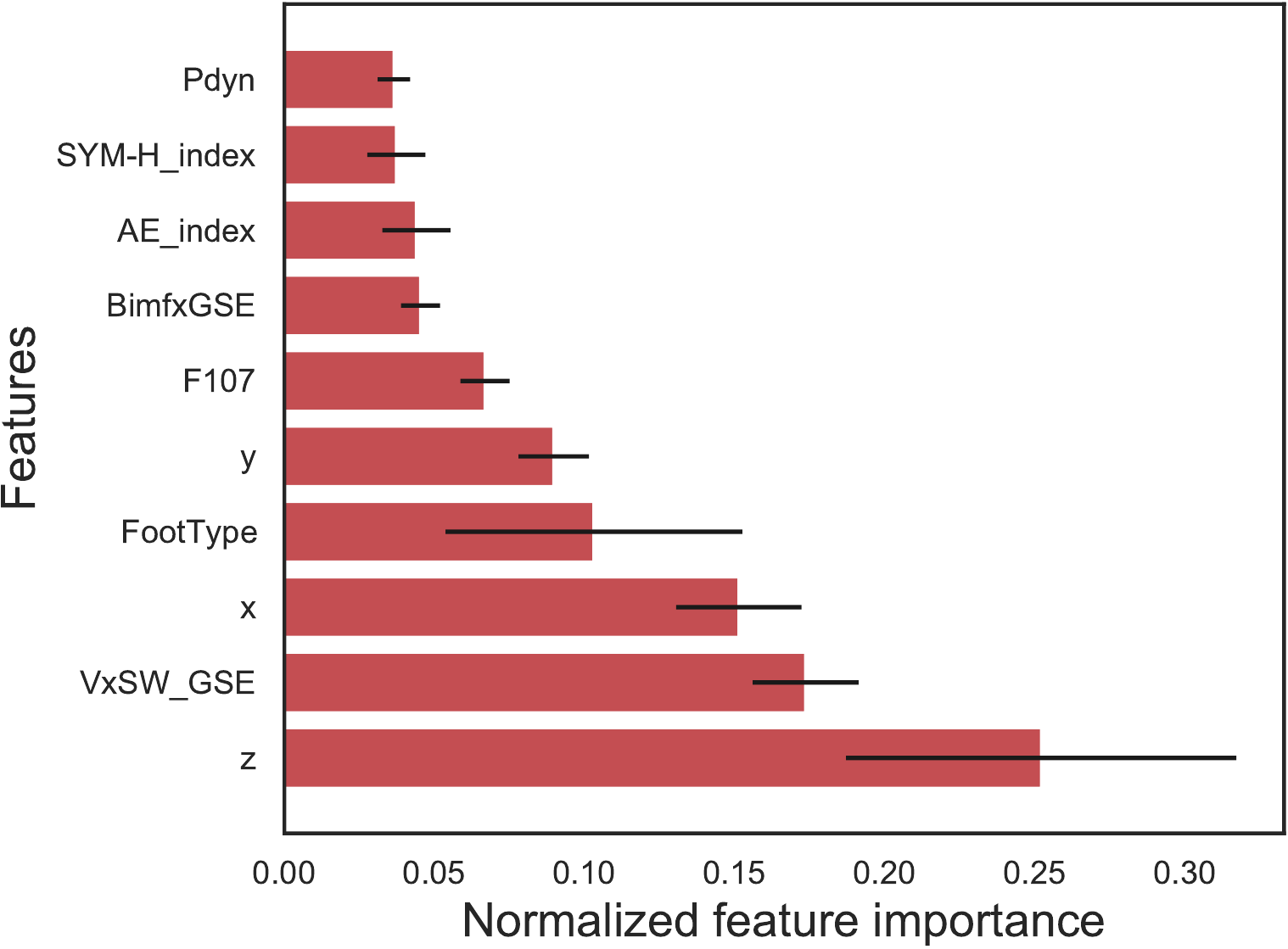}
	\caption{Importances of the different parameters in prediction of the contaminating count rates based on training data set. The black horizontal lines represent standard deviations.  \label{fig:importance}}
\end{figure}

The relative importances are stored  as an output in the fitted regression model. This is an array with shape corresponding to the number of features. The values of the array are positive and sum to 1.0. The higher the value, the more important is the contribution of the feature to the regression model. The relative importances of the features are plotted in Figure \ref{fig:importance}. The relative importances predicted by the Extra Trees Regressor algorithm are consistent with Pearson correlations in Figure \ref{fig:corr} and relations demonstrated in Figure \ref{fig:binned}: the location of the satellite, especially in $Z$ direction, the radial SW velocity and the Foot Type are the most important parameters for the prediction of the contamination count rates. 

We note that there are also other approaches to estimate feature importance such as Shapley values and permutation methods \citep{Shapley53,Breiman01}. Consideration of these methods, however, is beyond the scope of this work. The physics associated with important parameters is discussed in Section \ref{sec:discussion}.

\subsection{Testing the model}  \label{sec:results}

\begin{figure}[ht!]
	\plotone{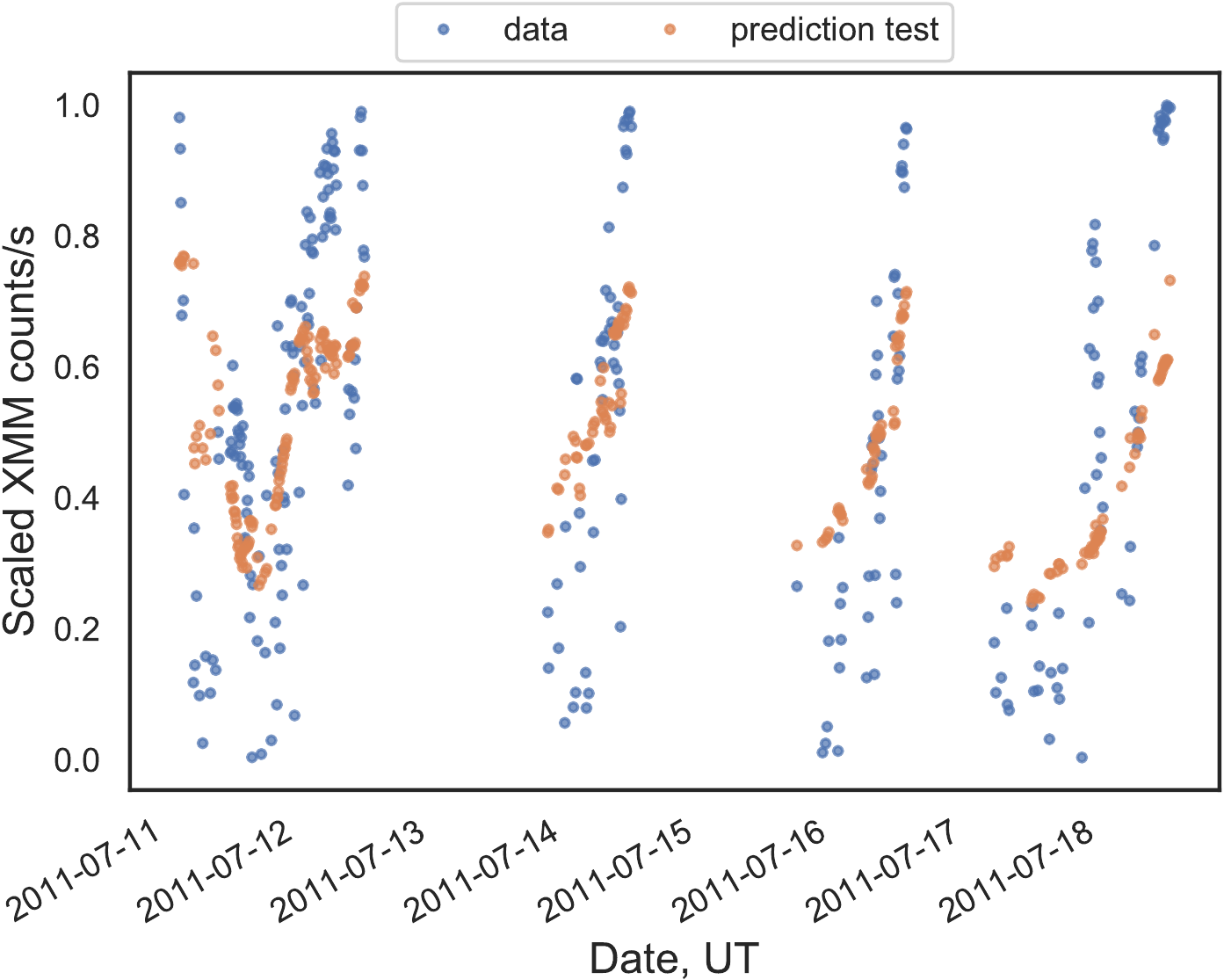}
	\caption{Time profiles of an interval from the test data set of observed data and those predicted by the ML model. \label{fig:test}}
\end{figure}

We test the model on the available data from 2011 to 2012. The diagram of the model performance is shown in Figure \ref{fig:trained} (right).  The distribution of the observed and predicted counts indicates that the model (on both test and trained data sets, see Figure \ref{fig:trained})  underestimates high values and overestimates low values. This is also seen well in Figure \ref{fig:test}, which presents the model performance on the test data set for a time interval in 2011. The performance of the trained model on the whole test data set, evaluated by different estimators, is listed in Table \ref{tab:performance}. The MSE error is close to zero. This is consistent with the ability of the model to predict mean values of count rates.  The $R^2$ indicates that the model can explain about 20\% variability of the count rates. The Pearson and Spearman correlation coefficients are moderate and they are statistically significant. To be significant at $t$=0.05 level, where $t$ is the Student coefficient, the Pearson coefficient, $r$ needs to exceed the value defined by the following expression: $$r=\frac{t}{n-2+t^2},$$ where $n$ is the number of samples \citep{Kendall73}. In our case taking $t$=2.8 for the two-sided distribution and n=7341, the number of values in the test data set, we get that values of $r>$0.03 will give significant correlation. Our values of $r$ are much higher, implying the significance of the model.

\subsection{Comparison with models based on best individual predictors}
\begin{figure}[ht!]
	\includegraphics[width=0.48\textwidth]{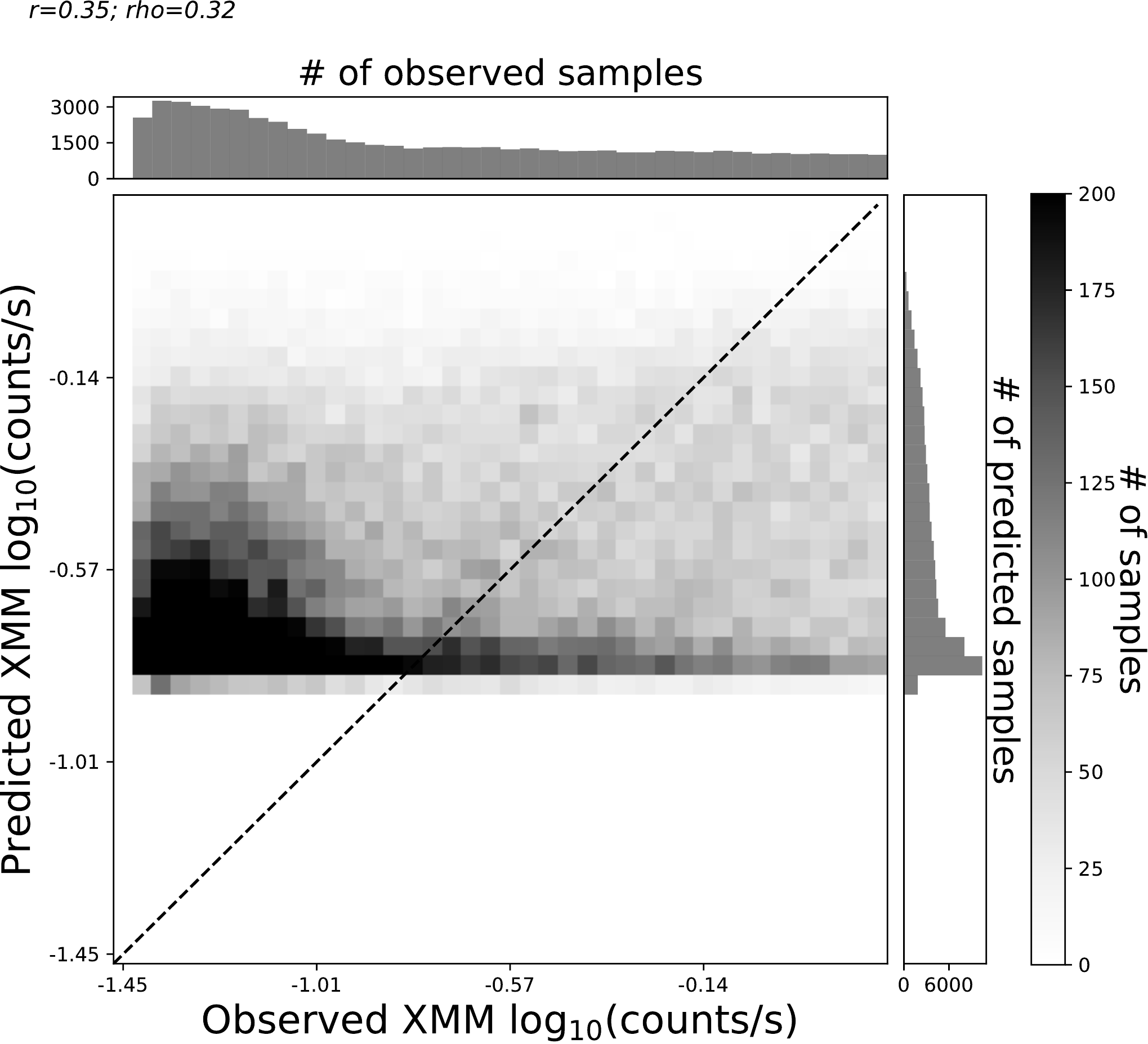}
        \hspace{4mm}
        \includegraphics[width=0.48\textwidth]{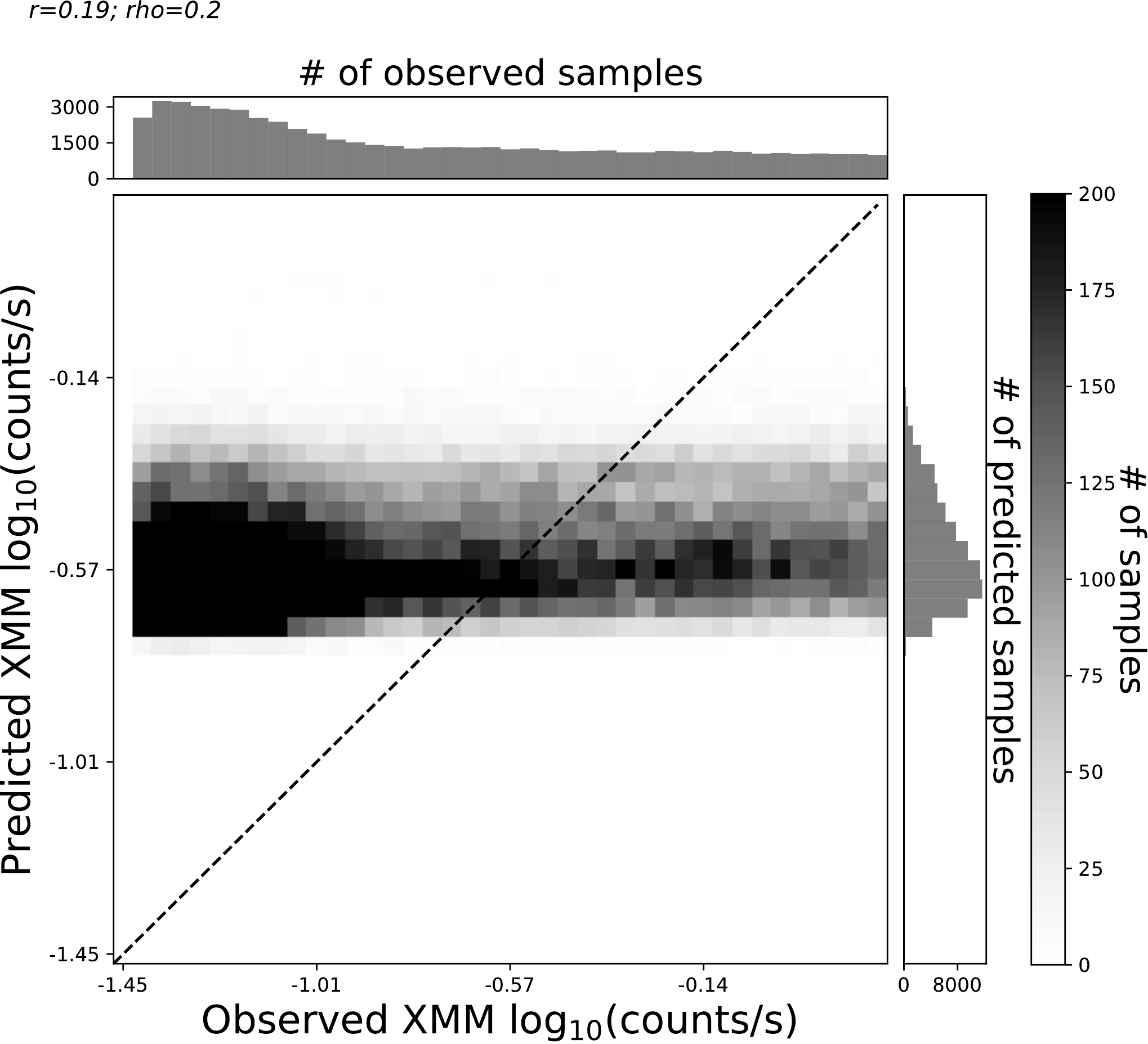}
	\caption{Observed count rates from the trained data set versus those predicted by the linear models: depending on ZGSE (left) and \textit{VxSW\_GSE} (right). The color represents number of samples.  \label{fig:linear}}
\end{figure}

Performance of the ML model is moderate and it is better than those of linear models derived with best individual predictors. In Figure \ref{fig:linear} one can see the performance of the linear models using the  training  data set. The models not only strongly underestimate and overestimate high and low values, respectively, but also have lower performance estimated by Pearson and Spearman correlations, which are 0.35 and 0.32 for the model on ZGSE and 0.19 and 0.2 for the model on \textit{VxSW\_GSE} and by other metrics, see Table \ref{tab:performance}. 

Less than optimal performance of the ML model can be explained by strongly non-linear behavior of the energetic charged particles that trigger the contamination on short time scales. The performance of our ML model can be improved in the future by adding more data.  The current data set covers just about 1 solar cycle and contains many data gaps (see Figure \ref{fig:test}). However, the complete magnetic cycle of the Sun spans two solar cycles. This could introduce further variations, that we do not cover.

\section{Discussion: delineating physics behind contamination}  \label{sec:discussion}
\subsection{Spatial dependencies of contamination}\label{sec:spatio} 

The dependence of the SP contamination count rate on the spacecraft position in GSE coordinates is in agreement with previous results, e.g., by \citet{Kuntz08, Ghizzardi:2017}.  Our results also agree with those from \citet{Walsh14} that show the strongest contamination observed on the closed field lines (see Figure \ref{fig:sketch}). The closed field lines are associated with the plasma sheet and the trapped particle population in the ring current and radiation belts. These are main reservoirs of energy in the magnetosphere. At higher latitudes, regions with open field lines or IMF (see Figure \ref{fig:sketch}) become more important, leading to the decrease in the ZGSE direction. These regions are typically associated with particle energies well below the SP range.  Indeed, SP count rates in the IMF are significantly lower than on closed field lines. Those on open magnetic field lines show the weakest SP count rates.  The plasma on the IMF can be accelerated by shock related processes discussed in Sections \ref{sec:HSS} and \ref{sec:shock}.   

The duskward asymmetry of the contamination can  partially  be explained by loss of energetic particles towards dawn side  because of different loss mechanisms, see \citep[e.g.][]{Kron14}. Such asymmetry is observed for energetic protons ($>$274 keV) and even stronger for energetic oxygen, see e.g. \cite{Kron15}. 

\subsubsection{Influence of IMF direction}

 The general direction of the solar wind Parker spiral \citep{Parker58} towards the Sun-Earth line, $\phi$ is $\sim$45$^{\circ}.$  In our data set, the  average IMF  components and confidence intervals are   \textit{BimfxGSE}= 0.018$\pm$0.025 nT, \textit{BimfyGSE}= -0.05$\pm$0.03 nT and \textit{BimfzGSE}= 0.072$\pm$0.023 nT. The average direction of the Parker spiral is $\phi\sim$44$^{\circ}.$ This geometry leads to formation of a quasi-parallel bow shock at the dawn side and a quasi-perpendicular bow shock at the dusk side. The quasi-parallel bow shocks are strong accelerators of plasma, see Section \ref{sec:shock}. Therefore, it would be expected to observe more contamination at the dawn side. However, this is not observed and consequently the direction of Parker spiral cannot explain the duskward asymmetry. 

At higher latitudes, the dayside and duskward flank preference of the contamination (see Figure (\ref{fig:binned}, c) for the dayside asymmetry and Figure \ref{fig:distr} for the duskward asymmetry) can be explained by the location of acceleration sources for particles at the dayside and the location of reconnection \citep[e.g.][]{Luo17}. In this study, the asymmetries in the spatial distributions of energetic ions were related to the location of the reconnection. In case of an average dawnward and northward IMF direction, the reconnection location is expected at dusk side  high latitudes  in the southern hemisphere \citep{Luo17}.  Particles on the reconnected field lines are further accelerated,  e.g.,  in a diamagnetic cavity (a region with low magnetic field formed during reconnection close to cusp (see sketch in Figure \ref{fig:sketch}  and Figure 6 in \citet{Nykyri11}). This  region traps  and accelerates plasma  particle population  that then  can penetrate inside the magnetosphere  and populate it  \citep[e.g.][]{Nykyri12,Sorathia19}.  In the data, the direction of the IMF is slightly northward and dawnward. However, the errors introduced by the processing of the OMNI data are in the range of 0.2 nT \citep{Papitashvili05}. Therefore, we cannot statistically confirm this explanation. More work is needed in this direction.

\subsection{Dependence on SW velocity}

Significant growth of the logarithm of the count rates with increasing radial SW velocity, $V_x,$ means that $V_x$ can be considered as a most important space weather parameter related the XMM contamination. The increase of the SW speed leads to the compression of the magnetosphere. However, the count rates show rather weak dependence on the SW dynamic pressure, see Figure (\ref{fig:binned}, h). Therefore, additional processes associated with faster SW lead to enhanced contamination. 

\subsubsection{Influence of SW high speed streams}\label{sec:HSS}
High SW speed is  often  associated with the SW high speed streams (HSS). HSS mostly occur during declining phase of the solar cycle due to an increase in equatorial coronal holes, which are the source of HSS. In Figure (\ref{fig:binned}, i) one can see that the strongest contamination occurs during medium values of the parameter F10.7 which corresponds to the declining phase of the solar cycle. Regions in which HSS overtake slow SW are often associated with co-rotating interactions regions (CIRs). At larger heliospheric distances (beyond 1 Astronomical Unit (AU)), CIRs often form shocks, which accelerate ions \citep[e.g.][]{Richardson18}. These accelerated ions can then travel back towards the Sun, and have been observed within about 0.3 AU \citep[e.g.][]{Allen20}. Both the abundance and the composition of suprathermal ions associated with CIRs has also been shown to have a solar cycle dependence \citep[e.g.][]{Allen19}. These accelerated ions can enter into the magnetosphere via reconnection.

\subsubsection{Influence of SW-magnetosphere energy coupling}
The SW speed is proportional to the SW electric field that controls the magnetic reconnection rate at the dayside \citep{Dorelli19}. Increased SW speed will lead to increased reconnection rate. Most of SW-magnetosphere energy coupling functions are proportional to the SW speed  \citep[e.g.][]{Gonzalez94,Milan12,Wang14}.  The increase in the SW speed leads to more effective further transport of the reconnected magnetic field lines towards the tail and then, again via reconnection  on the night side,  back to the Earth  to complete the cycle. A brief disturbance ($\sim$3 hours) that causes energy release from the tail into the high latitude ionosphere is called a substorm. This will lead to deviation of the magnetic field on the ground in the  high-latitude  regions and will be reflected in the AE index. This agrees well with the AE index being in the set of parameters leading to better prediction of the ML model, see Figure (\ref{fig:importance}, g). Significant growth of the count rates with AE index at least up to 100 nT is observed. Substorm activity leads to strong acceleration of ions by processes associated with magnetic reconnection such as magnetic field dipolarization, see e.g., \citet{Grigorenko17}. Stronger substorm activity does not lead to more effective acceleration of ions (see the same result in \citet{Luo14}). This is probably related to more effective loss mechanisms of particles producing SP during high magnetospheric activity. For example, acceleration to higher energy can lead to  a decrease in the  SP population. This is a topic for future investigations. We also would like to note that AE index is measured in the northern hemisphere, although XMM observations are in the southern hemisphere. This fact may reduce correlation between observations of count rates and northern geomagnetic activity. \citet{Weygand08} have shown that observations of southern auroral region ground magnetometers are not always consistent with AE index.  

Increased SW speed on longer time scales (hours), e.g. during CMEs, may lead to  geomagnetic  storms. The SP count rates increase with decrease of the SYM-H index from 0 up to approximately -50 nT, see Figure (\ref{fig:binned}, l). At stronger magnetic storms non-linear behavior is observed. The same as strong substorms, strong magnetic storms can be associated with higher losses of SP. Such non-linear behavior of SYM-H index and AE index indicates that alone they are not necessary good parameters for prediction of the XMM contamination during geomagnetically active times. 

\subsubsection{Role of quasi-parallel bow shock}\label{sec:shock}
In Figure (\ref{fig:binned}, j) one can see an increase of the contamination during large absolute values of the IMF $B_x$ component. Large absolute values of the IMF $B_x$ will increase probability of the formation of the quasi-parallel bow shock (normal to the shock is parallel to the IMF direction), at least at the dayside magnetosphere. The quasi-parallel bow shocks are strong accelerators of plasma \citep[e.g.][]{Blandford78,Sundberg_2016,Kron09}. The shocks with higher Mach numbers, associated with higher solar wind speeds, lead to more effective ion acceleration \citep{Treumann09}. 


\subsection{Oxygen ions}
We compare the dependence of the SP count rates on the AE index and the SW dynamic pressure with the dependencies of proton and oxygen ions at 10 keV and $>$274 keV in the terrestrial plasma sheet observed by \cite{Kron12}. One can note that the trends in Figures (\ref{fig:binned}, h and k) are similar to the dependence of energetic hydrogen and oxygen ions ($>$274 keV) on the AE index and the SW dynamic pressure in that study. This  is consistent with the idea  that energetic protons at several hundreds of keV may produce contamination. Additionally, this indicates that oxygen ions may also contaminate the XMM telescope. \citet{Kron12} shows that the intensity of oxygen ions can be comparable with those of protons during disturbed magnetospheric activity.

\section{Conclusions and open questions}  \label{sec:conclusions}
In this paper we delineate which geometric, solar, SW, and geomagnetic parameters are related to strong contamination in the XMM telescope, derive prediction models and discuss the possible physical interpretation suggested by this approach. 
\begin{enumerate}
	\item We reveal strong association of the contamination with:
	(a) location of the satellite and, therefore, the region in space (the strongest and clear exponential dependence is derived for the southward direction, $Z$); (b) the radial SW speed (exponential dependence is derived) and
		(c) magnetic field line Foot Type (the strongest contamination is observed on closed field lines).
 \item We derived a model to predict contamination which utilizes an ensemble of predictors (Extra Trees Regressor). It shows better performance than models based on individual parameters such as $Z$ or $V_x.$ It also helps to quantify importance for non-linear relations.
 \item The analysis of relative importances of the parameters indicates that
(a) processes of acceleration related to formation of the quasi-parallel shock may play an important role in formation of the contaminating population. The indications for these are (i) relatively strong contamination at large absolute values of IMF $B_x,$ (ii) strong dependence on the SW velocity and
(iii)  stronger contamination at the dayside;
(b) acceleration processes associated with reconnection at the day side may also play an important role and
(c) SYM-H index and AE index alone are not necessary good parameters for prediction of the XMM contamination during geomagnetically active times. 
\item Similarity of the dependencies of the SP count rates and the energetic oxygen ($>$274 keV) in the plasma sheet on the AE index and the SW dynamic pressure gives a hint that oxygen may also contaminate XMM telescope.
       \end{enumerate}
Road map for future missions: 
	(a)  it is advisable to  avoid observations during  times associated with high solar wind speed in the near-Earth magnetospheric region and
        (b) the same is recommended for closed magnetic field-lines,  especially at the dusk flank in the southern hemisphere (asymmetries in the northern hemisphere are not studied here). 

In our next studies we will focus on the following questions:
(1) Which processes associated with the strong SW speed are effective accelerators of energetic particles? In particular, acceleration sources associated with reconnection at the day side (such as diamagnetic cavities at cusps) and quasi-parallel bow shocks require enhanced attention. (2) Which energy range of particles are most efficient at producing this contamination? (3) Are there losses of SP contaminating particles during high magnetospheric activities? (3) What role do energetic oxygen  ions  play in the contamination observed by XMM?
 We will address all these questions in future work. For this we plan to compare XMM observations with energetic particle observations by  the  Cluster mission.

\acknowledgments

We acknowledge use of NASA/GSFC's Space Physics Data Facility's OMNIWeb service and OMNI data. We acknowledge XMM data archive \url{https://www.cosmos.esa.int/web/xmm-newton/xsa}.  The AE and SYM-H indices used were provided by the WDC for Geomagnetism, Kyoto (http://wdc.kugi.kyoto-u.ac.jp/wdc/Sec3.html). F10.7 index can be found at https://spaceweather.gc.ca/solarflux/sx-5-en.php.  This work was conceived within the team led by Fabio Gastaldello on ``Soft Protons in the Magnetosphere focused by X-ray
Telescopes'' at the International Space Science Institute in Bern, Switzerland. NS is supported by NASA Earth and Space Science Grant 80NSSC17K0433. EK  is supported by German Research Foundation (DFG) under number KR 4375/2-1 within SPP ``Dynamic Earth''. We are thankful to Irina Zhelavskaya for advice related to ML.

\software{sklearn \citep{scikit-learn},
       scipy  \citep{2020SciPy-NMeth},
          numpy  \citep{vanderWalt11},
          pandas  \citep{mckinney-proc-scipy-2010},
          Matplotlib   \citep{Hunter07}, 
          mysql.connector
          }

\appendix

 \begin{figure}[ht!]
 	\plotone{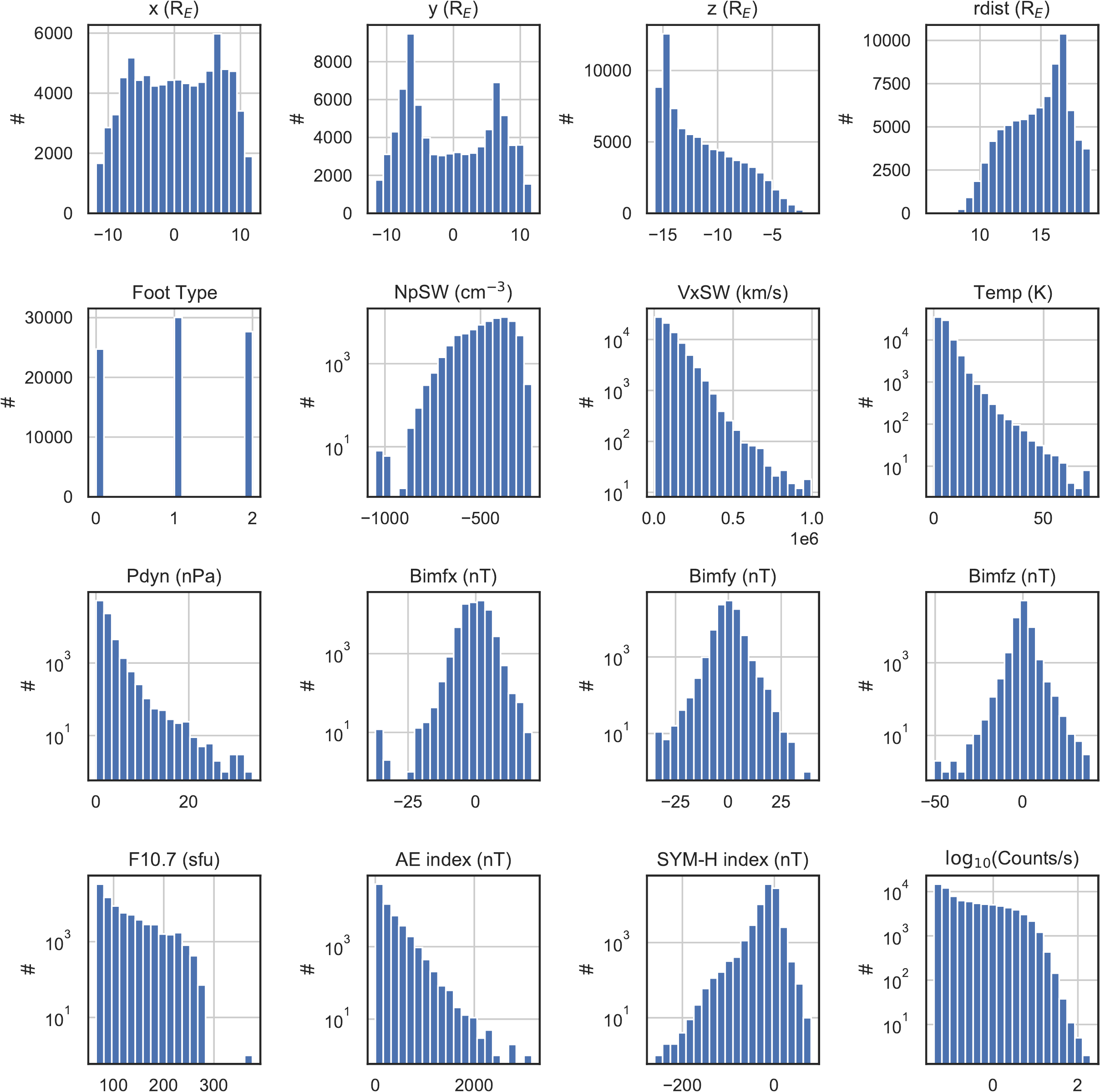}
 	\caption{  Histograms of the number of samples of  predictors and SP count rates. \label{fig:hist}}
 \end{figure}

\end{document}